\documentclass[superscriptaddress,aps,prd,twocolumn,showpacs,preprintnumbers]{revtex4}
\usepackage{graphicx}
\usepackage{amsmath,amssymb}

\begin{document}

\title{Warming up brane-antibrane inflation}

\author{Mar Bastero-Gil}
\email{mbg@ugr.es}
\affiliation{Departamento de F\'{\i}sica Te\'orica y del Cosmos, Universidad de Granada, Granada-18071, Spain}

\author{Arjun Berera}
\email{ab@ph.ed.ac.uk} 
\affiliation{SUPA, School of Physics and Astronomy, University of Edinburgh, Edinburgh, EH9 3JZ, United Kingdom}

\author{Jo\~ao G. Rosa}
\email{joao.rosa@ed.ac.uk} 
\affiliation{SUPA, School of Physics and Astronomy, University of Edinburgh, Edinburgh, EH9 3JZ, United Kingdom} 

\date{\today}

\begin{abstract}
We show that, in constructions with additional intersecting D-branes, brane-antibrane inflation may naturally occur in a warm regime, such that strong dissipative effects damp the inflaton's motion, greatly alleviating the associated $\eta$-problem. We illustrate this for D3-$\mathrm{\overline{D3}}$ inflation in flat space with additional flavor D7-branes, where for a Coulomb-like or quadratic hybrid potential a sufficient number of e-folds may be obtained for perturbative couplings and $\mathcal{O}(10-10^4)$ branes. This is in clear contrast with the corresponding cold scenarios, thus setting the stage for more realistic constructions within fully stabilized compactifications. Such models generically predict a negligible amount of tensor perturbations and non-Gaussianity $f_{NL}\sim\mathcal{O}(10)$.
\end{abstract}

 \preprint{Edinburgh 2011/14}

\maketitle

\section{Introduction}

While extremely successful as a phenomenological model, cosmological inflation \cite{inflation} still lacks a microphysical description within a fundamental theory of quantum gravity, a crucial feature given its large sensitivity to physical effects close to the Planck scale. 

This has motivated the search for inflaton candidates amongst the plethora of scalar fields present in string theory compactifications (see e.g.~\cite{Baumann:2009ni}). Brane-antibrane inflation \cite{Burgess:2001fx, brane_inflation} is one of the most attractive scenarios in this context, given the geometrical origin of the inflaton as the interbrane separation and the possibility of computing its potential for a broad class of background spacetimes. 

Inflation, however, does not seem to be generic in these systems. In flat space, for example, the interbrane separation required for achieving a sufficient number of e-folds exceeds the average size of the compact space \cite{Burgess:2001fx}. In more realistic scenarios, brane interactions are considerably weakened by the strong warping produced by fluxes and other nonperturbative effects required to stabilize the extra-dimensional moduli \cite{moduli_stabilization, Kachru:2003sx}. However, this typically leads to Hubble scale inflaton masses, a symptom of the generic $\eta$-problem in supergravity models with F-term SUSY breaking. The inflaton potential receives nevertheless a variety of additional contributions from SUSY breaking effects in the overall compactification that may cancel this large mass and improve its flatness, although typically at the expense of some fine-tuning \cite{warped_brane_potentials}. 

Warm inflation scenarios are generically free of the $\eta$-problem \cite{Berera:1995ie, BasteroGil:2009ec}, with significant dissipative effects slowing down the inflaton's motion and leading to the production of radiation during inflation, in many cases avoiding the need to reheat the system \cite{Berera:1995ie}. Dissipation may be achieved in a two-stage mechanism, with the inflaton coupling to heavy fields that in turn decay into light degrees of freedom, sustaining a nearly-thermal radiation bath at temperature $T>H$. This dissipation damps the inflaton's motion without spoiling the flatness of the potential through neither radiative nor thermal corrections if SUSY is softly broken and the temperature lies below the heavy mass threshold, with a sufficient number of e-folds requiring only moderately large multiplicities \cite{Berera:2002sp}.

In this work, we show that warm inflation scenarios can be easily implemented in more realistic D$\overline{\mathrm{D}}$-brane constructions with additional branes of different dimensionality intersecting both the brane and the antibrane stacks. This introduces light matter at the intersections that may be copiously produced during inflation, damping the D-brane motion, through interactions which are mediated by heavy strings stretching between the brane and antibrane stacks, naturally implementing the two-stage dissipation mechanism. We illustrate this for a D3-$\overline{\mathrm{D3}}$ system with additional D7-branes, which has been extensively studied in the context of the AdS/CFT correspondence (see e.g.~\cite{Erdmenger:2007cm}). A particular feature of these systems is the large field multiplicity that may be present during inflation, as brane-antibrane annihilation will naturally reduce the degrees of freedom to, for example, those found in the MSSM or its grand unified theory (GUT) extensions.

Although we restrict our analysis to flat space, where dissipation effects can be easily computed, we consider the potentials for which the $\eta$-problem is more severe, with additional effects typically flattening the potential, thus showing that warm inflation gives a robust solution to the $\eta$-problem in brane-antibrane inflationary scenarios. Furthermore, this constitutes the first explicit implementation of warm inflation in the context of a UV-complete theory \cite{warm_string, additional_fields}.

This article is organized as follows. In the next section, we describe the basic dynamical mechanisms leading to warm inflation scenarios, outlining the main features of the two-stage dissipation mechanism. In Sec.~III, we describe the inflationary dynamics in D3-$\overline{\mathrm{D3}}$ systems, in particular discussing the effects of additional D7-branes. We describe the associated field content and interactions in Sec.~IV, where the corresponding dissipative coefficient is also computed. In Sec.~V we present our main results for warm brane-antibrane inflation in flat space. Finally, in Sec.~VI, we discuss possible extensions to more realistic constructions and summarize the main conclusions of this work.

\section{Warm inflation dynamics}

Warm inflation scenarios \cite{Berera:1995ie} revisit an old proposal by L.~Z.~Fang \cite{Fang:1980wi} that particle production may have significant effects in inflationary expansion. As later proposed by Moss \cite{Moss:1985wn} and Yokoyama and Maeda \cite{Yokoyama:1987an}, this may be effectively implemented by introducing a friction term in the classical evolution of the inflaton field, $\phi$:
\begin{equation} \label{warm_inflation_eq}
\ddot\phi+3H\dot\phi+V_\phi=-\Upsilon\dot\phi~,
\end{equation}
where $H$ denotes the Hubble parameter and $V_\phi$ the first derivative of the inflationary potential with respect to the inflaton field. Encoding the interactions between the inflaton and light particles that may be produced and subsequently thermalized, such term acts as a source of radiation production:
\begin{equation} \label{radiation_eq}
 \dot\rho_R+4H\rho_R=\Upsilon\dot\phi^2~.
\end{equation}
While inflation may only occur when radiation gives a subdominant contribution to the energy balance of the Universe, i.e. $\rho_R\ll
\rho_\phi$, its effects on the inflationary dynamics may be significant when $\rho_R^{1/4}>H$, which roughly translates into $T>H$, assuming that thermalization occurs. In conventional or {\it supercooled} scenarios, interactions are assumed to be sufficiently feeble so that the friction term only becomes significant after slow-roll inflation, as the inflaton field starts to oscillate quickly about the minimum of its potential, with particle production reheating the Universe and providing a rapid transition to a radiation-dominated era. In warm inflationary scenarios, on the other hand, such a term may concur with Hubble friction to permit slow-roll solutions and allow for a smooth transition to a radiation-dominated phase, offering solutions to both the $\eta$- and the graceful exit problems. In particular, in the slow-roll regime \cite{BasteroGil:2009ec}, 
\begin{eqnarray} \label{slow_roll_eq}
3H(1+Q)\dot\phi\simeq -V_\phi~,\qquad 
4\rho_R\simeq3Q\dot\phi^2~,
\end{eqnarray}
where $Q=\Upsilon/(3H)$, leading to the slow-roll conditions \cite{Moss:2008yb}:
\begin{eqnarray} \label{slow_roll_cond}
\epsilon_\phi,~\eta_\phi,~\beta_\Upsilon<1+Q~,
\end{eqnarray}
where besides the conventional slow-roll parameters, one introduces $\beta_\Upsilon=m_P^2(\Upsilon_\phi V_\phi/\Upsilon V)$ to account for the field dependence of the dissipative coefficient. Strong dissipation may then allow for a sufficiently large period of accelerated expansion with potentials that would otherwise be too steep. One must also ensure that radiation is produced at a rate that compensates for the inflationary redshift
\begin{equation} \label{slow_roll_2}
\bigg|{d\ln\Upsilon\over d\ln T}\bigg|<4~, 
\end{equation}
and also that thermal effects do not induce a large inflaton mass, which translates into the condition:
\begin{eqnarray} \label{slow_roll_3}
\delta={T V_{T\phi}\over V_\phi}<1~. 
\end{eqnarray}

In \cite{Berera:1995wh}, Berera and Fang further suggested that the dynamics of the inflaton should be governed by a Langevin equation including not only the dissipative term but also a noise force term that would drive inflaton fluctuations, specified uniquely by a fluctuation-dissipation theorem. The primordial spectrum seeding the large scale structure of the observed Universe is then dominated by {\it thermal} rather than vacuum fluctuations of the inflaton field, due to the interplay with the thermal bath at $T>H$, giving \cite{Berera:1995wh, Berera:1999ws, Hall:2003zp}
\begin{eqnarray} \label{spectrum}
P_{\cal R}^{1/2} \simeq \bigg({H\over2\pi}\bigg)\bigg({3H^2\over V_\phi}
\bigg)\bigg({T\over H}\bigg)^{1\over2}(1+Q)^{5/4}\simeq5\times10^{-5},
\end{eqnarray}
where all quantities are evaluated at horizon crossing.

The microscopical basis for warm inflation has been the focus of many studies in the literature, being first examined by Berera, Gleiser and Ramos \cite{Berera:1998gx}, which set the adiabatic approach to dissipation within quantum field theory that has been used in all subsequent studies. Most models initially proposed considered all fields to be in a high-temperature phase, which raised some concerns on the models' possible realizations within quantum field theory in \cite{Berera:1998gx} and by Yokoyama and Linde \cite{Yokoyama:1998ju}, namely due to thermal effects spoiling the required flatness of the inflaton potential. 

Successful models have, however, been constructed based on the two-stage dissipation mechanism proposed by Berera and Ramos \cite{Berera:2002sp}, in which the inflaton is coupled to heavy mediator bosonic and fermionic fields, much heavier than the temperature of the Universe during inflation, which could in turn decay into light degrees of freedom. The production of the heavy particles is thus Boltzmann-suppressed, so that the particles are effectively in a zero-temperature state. In supersymmetric models, the leading quantum corrections will then cancel and hence maintain the flatness of the classical inflaton potential. Supersymmetry is, however, ineffective in cancelling time nonlocal processes, thereby allowing for strong dissipative effects while suppressing dangerous radiative and thermal corrections. A generic superpotential realizing this two-stage mechanism is given by \cite{BasteroGil:2009ec}
\begin{equation} \label{warm_superpotential}
 W=g\Phi X^2+hXY^2~,
\end{equation}
where the inflaton corresponds to the scalar component of the superfield $\Phi$, while the heavy mediators and the light radiation degrees of freedom are encoded in the bosonic and fermionic components of the superfields $X$ and $Y$, respectively. The first term in Eq. (\ref{warm_superpotential}) then leads to scalar masses $m_X^2=2g^2\phi^2$, discarding the mass shifts induced by soft SUSY breaking during inflation, and couples the inflaton fluctuations to the heavy mediators. The cross term in $|F_X|^2$  then allows for the decay of the heavy fields into the light degrees of freedom. Note that this precludes a direct decay of the inflaton into the light fields, which is crucial in keeping radiative corrections under control. Thermal corrections in this setup were computed in \cite{Hall:2004zr} and shown to give subleading contributions to the inflaton potential. The interactions arising from this type of superpotential yield, in the low-temperature regime $T<m_X/10$, a leading dissipative coefficient of the form $\Upsilon=C_\phi T^3/\phi^2$, with the constant parameter enhanced by the multiplicity of the heavy mediators and number of available decay channels, $C_\phi\simeq 0.16 h^4 \mathcal{N}_\mathrm{X}\mathcal{N}^2_\mathrm{decay}$ \cite{dissipation_coefficient}.

\section{Brane-antibrane inflation}

The simplest example of brane-antibrane inflation in type IIB string theory corresponds to two stacks of $N_c$ D3-branes and $\tilde{N}_c$ $\overline{\mathrm{D3}}$-branes spanning the noncompact directions 0123 and separated by a distance $r$ in a six-dimensional flat torus of volume $V_6=L^6$. At large distances, the two stacks interact mainly via the exchange of bulk massless modes, yielding a Coulomb-like potential that can be computed e.g.~by treating the antibrane stack as a probe of the geometry produced by the D3-branes at the origin of the compact space, giving to leading order:
\begin{eqnarray} \label{inflaton_potential}
V(\phi)=V_0\bigg[1+{\gamma_n\over n}\bigg({\phi\over m_P}\bigg)^{n}\bigg]~,
\end{eqnarray}
with $n=-4$ and
\begin{eqnarray} \label{inflaton_parameters}
V_0={g_s^3\tilde{N_c}\over 4\pi}\bigg({2\pi l_s\over L}\bigg)^{12}m_P^4~,\qquad
\gamma_{-4}={N_c\tilde{N}_c\over \pi^2}{V_0\over m_P^4}~,
\end{eqnarray}
where $g_s<1$ is the string coupling, $l_s$ is the fundamental string length, $m_P$ is the reduced Planck mass and $\phi=\sqrt{T_3\tilde{N}_c}r$ is the inflaton field, with $T_3=(g_s(2\pi)^3l_s^4)^{-1}$ denoting the D3-brane tension. The $\overline{\mathrm{D3}}$-branes' backreaction may be neglected if $\tilde{N}_c\ll N_c$ for slowly moving branes \cite{Easson:2007dh}, while the supergravity approximation holds for weak curvatures, $r\gg R$, with the near-horizon curvature scale $R/l_s=(4\pi g_s N_c)^{1/4}\gtrsim1$ in the regime of interest to our discussion \cite{Burgess:2003qv}.

This potential is characterized by a slow-roll parameter
\begin{equation} \label{eta}
\eta=-{10\over\pi^3}N_c\bigg({L\over r}\bigg)>1~,
\end{equation}
so that a sufficient number of e-folds cannot be obtained in a supercooled scenario, given that $r<L/2$ in a compact space. Furthermore, the $\overline{\mathrm{D}3}$-brane's motion is conformally coupled \cite{Kachru:2003sx}, yielding a mass term $H^2\phi^2$ that may dominate over the Coulomb term. This leads to a hybrid potential with $n=2$ and $\gamma_2=\eta=2/3$, also precluding a large number of e-folds in a particular manifestation of the generic supergravity $\eta$-problem.

As described in the previous section, successful realizations of warm inflation that could overcome the steepness of the inflaton potential require the existence of heavy mediatiors that catalyze the dissipative processes. This simple setup in fact includes heavy states coupled to the inflaton field, corresponding to massive strings stretched between the D3- and $\overline{\mathrm{D3}}$-brane stacks and the mass of which is proportional to the interbrane distance. There are, however, no light matter particles into which these states may decay, being the lightest states charged under the gauge group of both branes. 

The existence of such light states is nevertheless required if one aims to incorporate the Standard Model or any of its extensions in the D3-branes that survive annihilation, once the brane-antibrane tachyonic instability \cite{Sen:1998sm} ends inflation in a hybridlike fashion \cite{Burgess:2001fx}. This can be implemented, for example, by placing the stack of D3-branes at orbifold singularities and/or by including additional branes of different dimensionality \cite{Burgess:2004kv}.

A simple extension of the D3-$\overline{\mathrm{D3}}$ inflationary scenario is given by the inclusion of $N_f$ D7-branes wrapping e.g.~the 4567 compact directions, which introduces light matter particles corresponding to open strings attached to the latter and to each of the two D3-stacks. Before analyzing the field content and interactions in this extended setup and how these may lead to realizations of warm inflation, one should first examine how the inflaton trajectory is modified in this setup.

D7-branes source the type IIB axio-dilaton field, so that one expects them to modify the constant dilaton profile of the D3-brane geometry. In fact, supergravity solutions combining D3- and D7-branes have been shown to exhibit a logarithmic dilaton profile near the D7-branes' horizon, yielding a local string coupling \cite{Kirsch:2005uy}:  
\begin{eqnarray} \label{string_coupling}
g(\rho)=g_s\bigg[1-{g_sN_f\over4\pi}\log\bigg({(\rho-d)^2\over d^2}\bigg)\bigg]^{-1}~,
\end{eqnarray}
where $\rho$ denotes the radial direction in the plane transverse to the D7-branes, with the D3-branes located at $\rho=0$, with $g(0)=g_s$, and the D7-branes at $\rho=d$, where the coupling vanishes. This modifies the potential energy of a mobile $\overline{\mathrm{D3}}$-brane probing this geometry. In particular, as the brane tension is inversely proportional to the local string coupling, when placed randomly in the compact dimensions it will quickly move towards the origin along the $\rho$ direction in order to minimize its energy. 

The presence of the D7-branes thus selects a particular inflationary trajectory with both D3- and $\overline{\mathrm{D3}}$-brane stacks separated along the D7-branes' worldvolume at a distance $d$ from the latter. The potential is nevertheless given by Eq. (\ref{inflaton_potential}), with the inflaton field now corresponding to the interbrane separation $r$ along the compact directions wrapped by the D7-branes.


\section{Two-stage dissipation with D-branes}

Let us first consider a setup with two concident D3-branes and $N_f$ D7-branes. This yields an $\mathcal{N}=2$ worldvolume theory, described by a $U(2)$ $\mathcal{N}=4$ vector multiplet and an $\mathcal{N}=2$ hypermultiplet. In $\mathcal{N}=1$ language, these include adjoint chiral multiplets $\Phi_i$, $i=1,2,3$, and $N_f$ (anti)fundamental chiral superfields $Q,\tilde{Q}$ corresponding to oriented open strings at the D3-D7 intersections, with a superpotential \cite{Erdmenger:2007cm}
\begin{eqnarray} \label{superpotential}
W=\sqrt2g_{YM}\mathrm{Tr}\Phi_1[\Phi_2,\Phi_3]+\sqrt2g_{YM}\tilde{Q}\Phi_3Q~,
\end{eqnarray}
where $\Phi_{1,2,3}$ correspond to the (45), (67) and (89) coordinates, respectively, and $g_{YM}^2=2\pi g_s$ \cite{Johnson}. Notice that, in particular, the scalar vacuum expectation value of $\Phi_3$ determines the D3-D7 separation, which justifies its coupling to the (anti)fundamental superfields. The gauge coupling on the D7-branes is suppressed by the volume of the compact cycle wrapped by the branes, so that D7-D7 strings will be irrelevant to our discussion. When the D3-branes, denoted by A and B, are separated along the compact directions, i.e.~by turning on a scalar vacuum expectation value for a linear combination of the $\Phi_i$ fields, the gauge theory is broken to $U(1)_A\times U(1)_B$, under which the superfields may be decomposed as
\begin{eqnarray} \label{decomposition}
\Phi_i=\left(
\begin{tabular}{cc}
$\Phi^A_i$ & $X_i$ \\
$\tilde{X_i}$ & $\Phi^{B}_i$ 
\end{tabular}\right)~,\ 
Q=\left(
\begin{tabular}{c}
$Y^A$  \\
$Y^B$ 
\end{tabular}\right)~,
\ 
\tilde{Q}=\left(
\begin{tabular}{c}
$\tilde{Y}^A$  \\
$\tilde{Y}^B$ 
\end{tabular}\right)~.
\end{eqnarray} 
In this notation, the superfields $\Phi^{A,B}_i$ correspond to the positions of each D3-brane in the compact space, while $X_i$ and $\tilde{X}_i$ correspond to both longitudinal and transverse excitations of the bifundamental strings stretched between the two branes, with both orientations. Similarly, the $Y^{A,B}$ and $\tilde{Y}^{A,B}$ superfields correspond to oriented open strings in the $D3^{A,B}-D7$ sectors, transforming in (anti)fundamental representations of the corresponding $U(1)$ gauge group. With this decomposition, the superpotential Eq. (\ref{superpotential}) can be written as:
\begin{eqnarray} \label{decomposed_superpotential}
W&=&\sqrt2 g_{YM}\bigg[\epsilon_{ijk}\Phi^{AB}_i X_j\tilde{X}_k+\tilde{Y}^AX_3Y^B+\tilde{Y}^B\tilde{X}_3Y^A\nonumber\\
&+&\tilde{Y}^A\Phi^{A}_3Y^A+\tilde{Y}^B\Phi^B_3Y^B\bigg]~,
\end{eqnarray} 
where $\Phi^{AB}_i=\Phi^A_i-\Phi^B_i$. For the particular inflationary trajectory described in the previous section, one may choose w.l.o.g. $\phi=\sqrt2\langle|\Phi_1^{AB}\rangle|\neq0$, such that the relevant terms in the superpotential Eq. (\ref{decomposed_superpotential}) are
\begin{eqnarray} \label{decomposed_superpotential_1}
W=\sqrt2 g_{YM}\big[\Phi^{AB}_1X_2\tilde{X}_3+\tilde{Y}^B\tilde{X}_3Y^A+\ldots\big]~,
\end{eqnarray} 
and similarly for $X_3$. The $F$-term scalar potential then yields $m_X=g_{YM}\phi=r/(2\pi l_s^2)$ for the $X_{2,3}$ and $\tilde{X}_{2,3}$ scalar states, which as expected corresponds to the mass of open strings stretched between the D3-branes. These strings also describe the massive vector bosons of the broken gauge symmetry, which acquire an equal mass by incorporating the linear combination of the $X_1$ and $\tilde{X}_1$ scalar components that corresponds to longitudinal excitations of the open strings. The orthogonal linear combination also obtains a mass $m_X$ from D-terms in the scalar potential, completing the massive vector multiplet resulting from the supersymmetric Higgs mechanism.

Notice that an inflationary scenario requires one of the D3-branes to be replaced with an antibrane so that supersymmetry is softly broken, which translates into shifting the $X_i$ and $\tilde{X}_i$ scalar masses in opposite directions, while keeping the corresponding fermion masses unchanged, with $\mathrm{Str}M^2=0$ \cite{Brodie:2001pw}. This shift is negligible for large brane separations and only becomes significant at the end of inflation as $r\sim l_s$. 

It is thus clear that the interactions in Eq. (\ref{decomposed_superpotential_1}) yield the two-stage dissipation mechanism described in Sec.~II, such that the inflaton is coupled to the heavy $X_2$ and $\tilde{X}_2$ fields, which may in turn decay into the $Y$ and $\tilde{Y}$ D3/D7 states. The latter couple to the $\Phi_3^{A,B}$ fields but not to the inflaton, and remain light for $d\ll r$. Field multiplicities are enhanced in the non-Abelian case, with $\mathcal{N}_X=2N_c\tilde{N}_c$ and $\mathcal{N}_{decay}=N_f$, thus allowing for strong dissipative effects. Notice that, even in the non-Abelian case, the inflaton direction does not have any self-interactions and only appears linearly in the superpotential, similarly to the GUT-inflation scenario of \cite{Dvali:1997mh}.

The associated dissipation coefficient in the low-temperature regime, $T\lesssim m_X/10$, is given by \cite{dissipation_coefficient}
\begin{equation}
C_\phi\simeq 131.4\ \alpha_{YM}^2N_c\tilde{N}_cN_f^2~.
\end{equation}
Notice that in this computation one may neglect the background curvature for $r\gg R$, given the large mass of the mediator fields during inflation, $m_XR=(r/R)(R/2\pi l_s)^2\gg1$. Also, higher-loop contributions, which can enhance dissipation, are suppressed by powers of $(T/m_X)^2$, despite the moderately large 't Hooft coupling, $g_s N_c\gtrsim 1$, required for strong dissipation. Although this precludes a perturbative treatment of the radiation fluid in both the D3 and $\overline{\mathrm{D3}}$ sectors, we expect interactions to be strong enough to sustain the required near-thermal equilibrium at $T>H$. 


\section{Results}

With the results of the previous sections, we may now analyze the dynamics of warm inflation in the D3/D7-brane construction. In the slow-roll approximation, the evolution equations for a potential of the form Eq. (\ref{inflaton_potential}) are given by 
\begin{eqnarray} \label{warm_equations}
{d\ln \hat\phi\over d N_e}&=&-{\gamma_n\over1+ Q}\hat\phi^{n-2}~,\nonumber\\
{d\ln Q\over d N_e} &=&{\gamma_n\over1+7Q}\hat\phi^{n-2}(14-6n+5\gamma_n
\hat\phi^n) ~,
\end{eqnarray}
where $\hat\phi=\phi/m_P$ and $N_e$ is the number of e-folds \cite{BasteroGil:2009ec}. We have solved these equations assuming either the Coulomb-like or the conformal mass term as dominant interactions in order to determine the number of branes and couplings required for warm inflation with $H<T<m_X/10$ and a subdominant radiation component, $\rho_R<V_0$. The interbrane distance is restricted to $l_s<r<L/2$ and we take $r\sim L/2$ at horizon crossing. The compactification size $L$ is determined by the amplitude of the primordial spectrum, given by Eq. (\ref{spectrum}). Also, the number of relativistic degrees of freedom produced by dissipative effects is given by 
\begin{equation}
g_*={15\over2}(N_c+\tilde{N}_c)N_f~.
\end{equation}

Our results are summarized in Figs.~1 and 2, showing the regions of the plane $(N_c,\tilde{N}_c)$, with $N_c>\tilde{N}_c$, allowing for $N_e=50$ e-folds of warm inflation for different values of $\alpha_{YM}$. In Fig.~1, we take the maximum number of D7-branes for which a full supergravity solution is known to exist, $N_f=12$ \cite{Kirsch:2005uy}, while in Fig.~2 we take $N_f=100$ to illustrate the effect of increasing the number of light fundamental flavors. Notice that a large number of fundamental fields lighter than the temperature during warm inflation may be compatible with low-energy phenomenology if the latter lies above the TeV scale.

\begin{figure}[htpb]
\centering\includegraphics[scale=0.36]{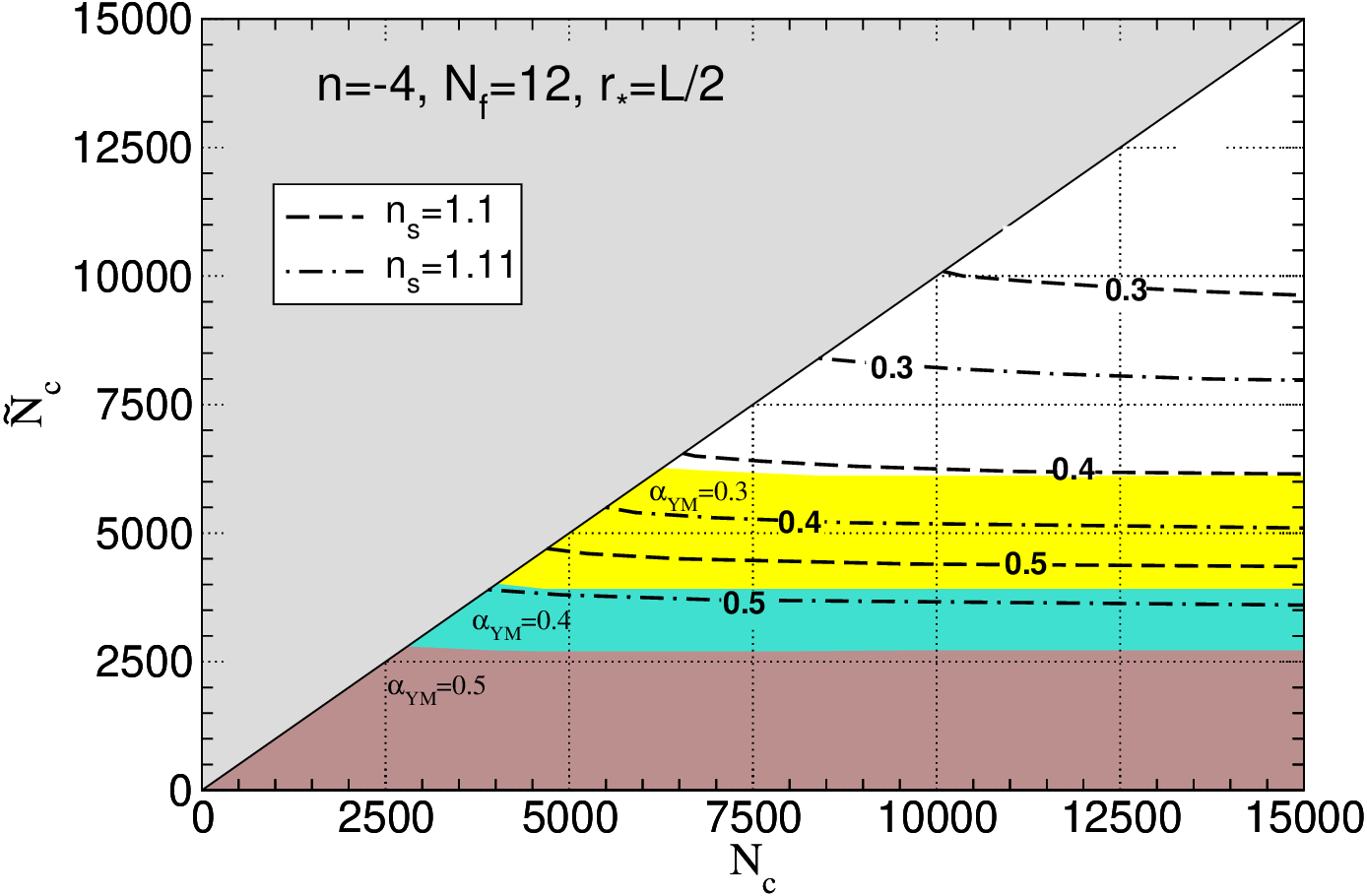}\vskip0.95cm
\centering\includegraphics[scale=0.36]{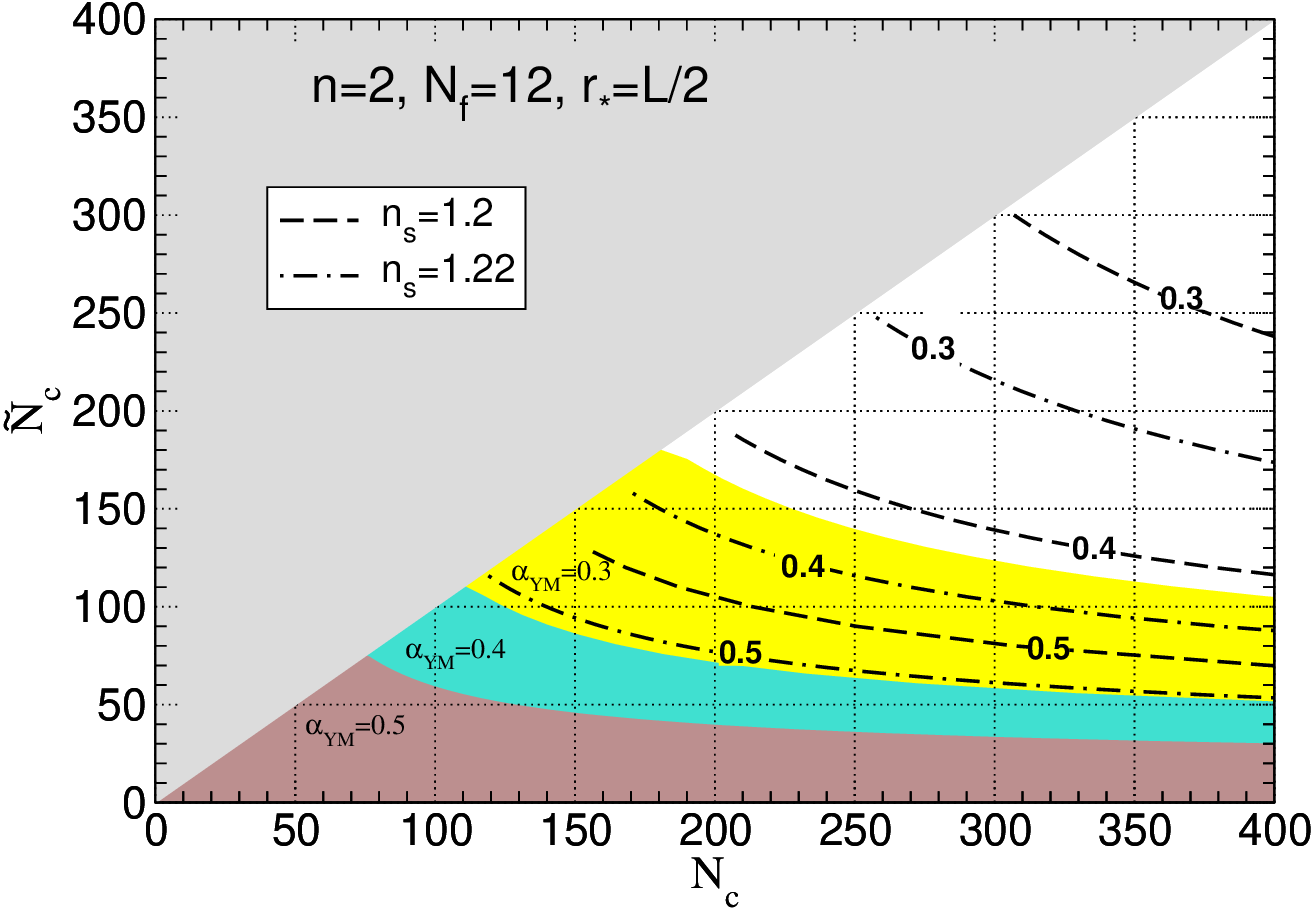}
\caption{Results for $N_e=50$ with the $n=-4$ (top) and $n=2$ (bottom) hybrid potentials, with $N_f=12$ light fields. In
the colored regions there are no solutions for $\alpha_{YM}$ below the given value. The associated value of $\alpha_{YM}$ is specified in each line of constant $n_S$.} 
\label{fig1}
\end{figure}

\begin{figure}[htpb] 
\centering\includegraphics[scale=0.36]{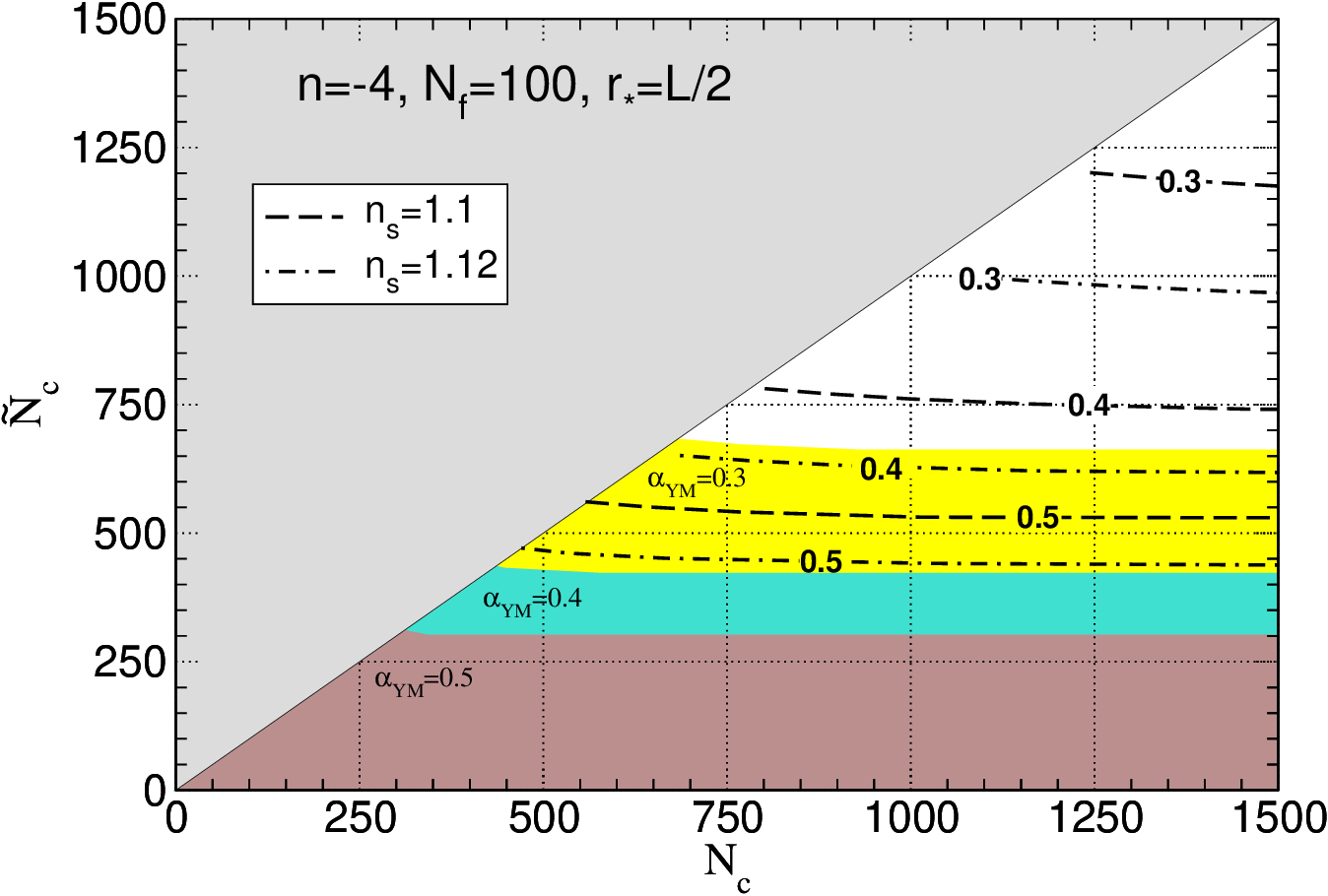}\vskip0.95cm
\centering\includegraphics[scale=0.36]{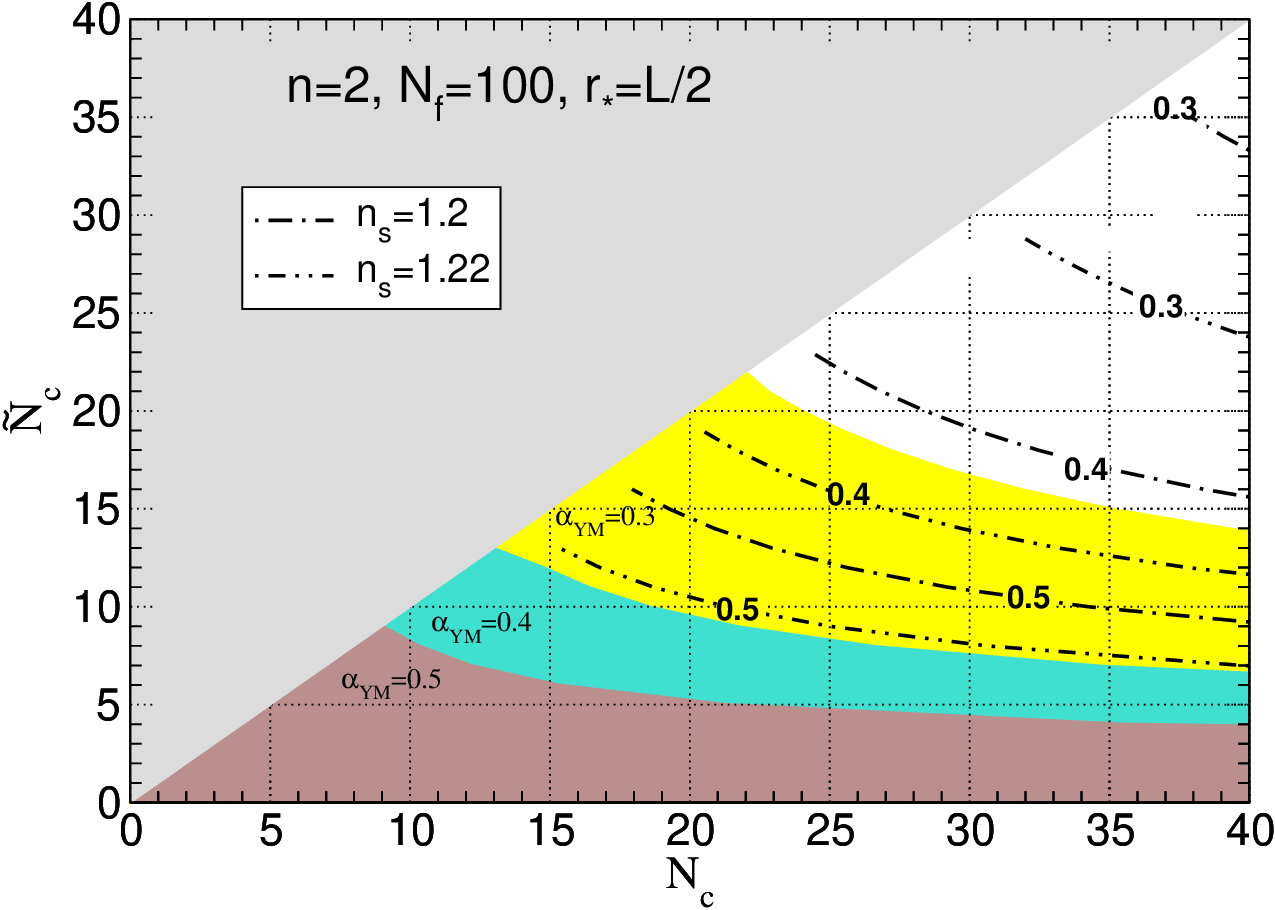}
\caption{Results for $N_e=50$ with the $n=-4$ (top) and $n=2$ (bottom) hybrid potentials, with $N_f=100$ light fields. In
the colored regions there are no solutions for $\alpha_{YM}$ below the given value. The associated value of $\alpha_{YM}$ is specified in each line of constant $n_S$.} 
\label{fig2}
\end{figure}

For $n=-4$, the fast increase of $Q$ may lead to a premature radiation-dominated era, so that we exclude regions where $\rho_R>V_0$ for $N_e<50$. For $n=2$, on the other hand, the milder growth of dissipative effects, generic for smaller $|n|$, makes radiation redshift towards the onset of the tachyon instability, so that the condition $r>l_s$ for 50 e-folds determines the allowed regions in this case. 

As shown in Fig.~\ref{fig1}, successful warm inflation with a Coulomb-like potential can be obtained with $\mathcal{O}(10^3-10^4)$ D3-branes, whereas for the quadratic potential $\mathcal{O}(100)$ branes are sufficient, given that the associated $\eta$-problem is significantly less severe in this case. As expected, for larger gauge couplings the minimum number of D3-branes decreases and, in the perturbative limit  $g_s=2\alpha_{YM}\approx1$, we may have $\tilde{N}_c<75$ (2700) for $n=2$ ($-4$). As illustrated in Fig.~2, increasing the number of D7-branes allows for warm inflation with a significantly smaller number of D3-branes and antibranes, in particular, with $N_f=100$ and $g_s=2\alpha_{YM}\approx1$ we can have 50 e-folds of inflation with $\tilde{N}_c\simeq 10$ (350) for $n=2$ (-4).  

Our results are quite insensitive to $N_c$, as the $\mathrm{\overline{D3}}$-branes set the scale of inflation, so that we
may have $N_c\gg\tilde{N}_c$ within the probe brane approximation. Also, if some other mechanism determines the amplitude
of primordial fluctuations, the quadratic (Coulomb) potential leads to a period of warm inflation with $T\gtrsim H$ for $N_c,\tilde{N}_c\gtrsim 11$ (24)  for $\alpha_{YM}= 0.5$  and $N_f=12$.

For the allowed regions of Figs.~1 and 2, the scale of inflation lies in the range $10^{13}-10^{14}$ GeV. Given the enhancement of thermal fluctuations for stronger dissipative effects, the primordial spectrum implies a lower inflationary scale for larger numbers of branes and couplings. This corresponds to radiation temperatures between $10^{10}-10^{13}$ GeV  and relatively large compact spaces, with $L/l_s\sim 100-300$. The conformal mass term yields lower temperatures, due to the smaller dissipation at horizon crossing, although generically $Q_*\gtrsim1$. 

In Figs.~1 and 2 we also plot in each case the lines of constant spectral index, which for a hybridlike potential of the form
Eq. (\ref{inflaton_potential}) is given by \cite{BasteroGil:2009ec} 
\begin{eqnarray} \label{spectral_index}
n_S-1=\begin{cases}
2\gamma_n\big({\phi_*\over m_P}\big)^{n-2}, &Q_*\ll1\\
-{3\gamma_n(n-7)\over7Q_*}\big({\phi_*\over m_P}\big)^{n-2}, &Q_*\gg1
\end{cases}~.
\end{eqnarray}
The primordial spectrum is thus always blue-tilted for $n<7$, although deviations from scale invariance are suppressed for strong dissipation, i.e. a larger number of D-branes and/or couplings, as $n_S-1\propto Q_*^{-1}$. While present observational data favors a red-tilted spectrum assuming a simple power law, an analysis including tensor modes and a running spectral index suggests a blue-tilt and a large negative running \cite{Komatsu:2010fb}, which is in fact a feature of the quadratic hybrid potential, although a positive running is obtained for $n=-4$. As generic in warm inflation models, tensor perturbations are negligible, while a large non-Gaussianity with $f_{NL}\sim\mathcal{O}(10)$ is obtained \cite{non_gaussianity}, within the detectable range of Planck \cite{planck}.

Notice that, close to the limits of the allowed regions, we find $r\sim R\gtrsim l_s$ near the end of inflation for $n=-4$, with $r\sim l_s$ for $n=2$ as mentioned earlier, introducing string theory  corrections to the supergravity potential during the last few e-folds that may be significant. Also, the finite temperature of the light fields translates into a black hole horizon in the D3-brane geometry at \cite{Johnson} 
\begin{eqnarray} \label{horizon_radius}
r_H=\pi R^2 T= R\kappa^{-{1\over4}}\bigg({\rho_R\over V_0}\bigg)^{1\over4}~,
\end{eqnarray}
where
\begin{equation} \label{kappa}
\kappa={1\over4}\bigg({N_f\over\tilde{N}_c}+{N_f\over N_c}\bigg)~.
\end{equation}
This implies that finite temperature corrections are negligible for the quadratic potential, as radiation remains subdominant during inflation and $r\gg r_H$. For the $n=-4$ case, one may have $r_H\gtrsim R$ depending on the ratio between the numbers of D7- and D3-branes, in which case these corrections may become significant towards the end of inflation, with the temperature approaching the Hagedorn limit. One should also notice that the leading quantum corrections are already included in the potential Eq. (\ref{inflaton_potential}), as the string annulus diagram can be interpreted both as a tree-level closed string exchange or a one-loop open string process. Although it would be interesting to study these corrections in further detail, they do not modify our main results.

\section{Summary and future challenges}

In this work, we have analyzed the dynamics of warm inflation models in the context of brane-antibrane scenarios. We have shown that a two-stage dissipative mechanism arises naturally in constructions including additional D-branes, a necessary feature of D-brane realizations of the standard model.  

In particular, we considered a system where the inflating D3-$\overline{\mathrm{D3}}$ stacks move parallel to the worldvolume of a D7-brane stack, such that their motion is damped by dissipative effects arising from the production of light string states in the D3-D7 and $\overline{\mathrm{D3}}$-D7 sectors, mediated by heavy D3-$\overline{\mathrm{D3}}$ strings. The latter are the only states coupling directly to the inflaton field, thus keeping both radiative and thermal corrections under control for temperatures below the heavy mass threshold. 

Our analysis shows that successful realizations of warm inflation can be obtained with perturbative couplings and $\mathcal{O}(10-10^4)$ D3-branes, depending on the potential and number of light flavors, leading to a sufficiently long period of accelerated expansion even for the steep potentials dominating brane-antibrane interactions in flat space. Furthermore, such scenarios generically predict interesting observational signatures such as detectable non-Gaussianity and a negligible tensor-to-scalar ratio. 

Vacuum stabilized warped throats generically yield more complicated potentials than those analyzed in this work, with a lower inflationary scale. However, in a warped background, the brane Coulomb-like interactions are significantly weakened and, moreover, the effects of compactification generically introduce repulsive terms $-a_{\Delta}(\phi/m_P)^\Delta$ \cite{Baumann:2009ni} that may alleviate the $\eta$-problem by generating an inflection point in the potential, although the associated geometrical parameters need to be finely tuned. This fine-tuning can then be ameliorated in warm inflation scenarios, as dissipation allows for steeper potentials. Note that repulsive terms also shift the spectral index towards red-tilted values, in particular bringing the values shown in Figs.~1 and 2 closer to scale invariance, and also increase the growth of dissipative effects for $\Delta>7/3$, according to Eq.~(\ref{warm_equations}). Even for a flat torus, the D3-brane ``images'' yield a leading repulsive quadratic ``jellium'' term \cite{Shandera:2003gx}, which reduces the conformal value of $\gamma_2$, while higher-order corrections lead to more interesting brane trajectories \cite{Rosa:2007dr}, namely at distances $r\sim L/2$ as considered earlier. We have, thus, analyzed worst case scenarios and the required number of branes and couplings can be significantly reduced with the inclusion of additional effects.

Our independent analysis of two generic contributions to the interbrane potential, involving both positive and negative powers of the inflaton field, also illustrates how the number of branes required for successful warm inflation varies significantly with the steepness of the potential, also determining whether radiation will come to dominate the energy density before brane-antibrane pairs annihilate. In particular, the mechanism described in this work can be easily generalized to scenarios with D-branes of different dimensionality, where the associated Coulomb-like brane-antibrane potentials are generically flatter than for D3-$\overline{\mathrm{D3}}$ pairs such that successful warm inflation may be obtained with a significantly smaller brane multiplicity. Also, our analysis focused on the effects of fundamental matter fields, whereas including adjoint fields could provide a better estimate of dissipative effects in these constructions, although this lies outside the scope of this work. 

On considering possible realizations of warm inflation in realistic string compactifications, one should check that finite temperature effects during inflation do not destabilize the moduli fields. As shown in \cite{Buchmuller:2004tz}, this may occur if thermal contributions to the moduli potential exceed the barrier, $V_b$, separating the stabilized local minimum from the decompactification limit, defining a critical temperature $T_c\sim V_b^{1/4}$. However, one must also require that the vacuum energy during inflation does not overcome this barrier, so that generically $T_c\gg H$ and warm inflation scenarios are possible within stabilized compactifications. It is also possible for other moduli fields to mediate additional dissipative effects if they couple to both the inflaton and light degrees of freedom and remain heavy during inflation, $m_{mod}\gtrsim T$. These effects will necessarily enhance the contribution of open string states computed in this work, thus further relaxing the $\eta$-problem. 

Although the D3/D7-brane setup is phenomenologically unrealistic, we expect its main qualitative features to be present in scenarios where chiral matter is obtained, for example, by placing D3-branes at singularities near the bottom of a warped throat \cite{Burgess:2004kv}, where D7-branes may also be embedded (see e.g.~\cite{Erdmenger:2007cm}). Notice that this may yield larger Yukawa couplings in Eq. (\ref{decomposed_superpotential}), depending on the local geometry probed by the D3-branes, possibly leading to stronger dissipative effects with fewer branes. 

In such scenarios, the MSSM or its GUT extensions may be implemented in the D3-branes surviving annihilation \cite{Kallosh:2004yh}, which typically requires $N_c-\tilde{N}_c\sim \mathcal{O}(10)$, with $N_c,\tilde{N}_c\sim \mathcal{O}(10-100)$ being natural, thus favoring realizations with a dominant quadratic potential, although cosmological models with a larger number of branes in the early Universe have been considered in the literature (see e.g. \cite{Alexander:2000xv, Piao:2003sc}). A moderately large number of fundamental flavors $N_f$ may also be phenomenologically consistent if their masses lie above the current experimentally accessible energy scales but still behave as relativistic matter during warm inflation. Although we have, for simplicity, considered single stacks of coincident branes and antibranes, this is an unnecessary fine-tuning of the initial D-brane configuration. One could alternatively envisage multistage scenarios with a few smaller D3-$\overline{\mathrm{D3}}$ stacks yielding smaller dissipative effects, such that each pair provides only a fraction of the total number of e-folds. This would, for example, relax the need for $\mathcal{O}(10^3-10^4)$ coincident D3-branes with a dominant Coulomb-like potential, although a smaller number of branes is nevertheless expected in more realistic scenarios.

As some amount of radiation is produced during inflation, warm scenarios may avoid the need for an efficient reheating through tachyon condensation, which may also be delayed by finite temperature effects \cite{tachyon_reheating}. Furthermore, the dissipative $X$ fields that become tachyonic at the end of inflation are naturally coupled to radiation and one expects a significant fraction of the inflaton's energy density to be transferred into the surviving branes, so that the reheating temperature is not necessarily the same as the temperature during warm inflation. 

Our analysis shows that dissipative effects are a natural feature of intersecting brane models that greatly alleviates the $\eta$-problem in brane-antibrane inflation. We expect this to significantly reduce the fine-tuning of geometrical parameters and initial conditions in stabilized compactifications, a topic we will explore in future work.


\acknowledgements

We thank Arjun Bagchi for fruitful discussions. MBG is partially supported by MICINN (FIS2010-17395) and ``Junta de Andaluc\'ia'' (FQM101). AB and JGR are supported by STFC.


\end{document}